# Effect of current on terahertz plasmons in AlGaN/GaN heterostructures

M. Dub[1,2*], P. Sai[1,2], Y. Ivonyak[1,2], P. Prystawko[2], W. Knap[1,2], S. Rumyantsev[2]

[1]CENTERA, CEZAMAT, Warsaw University of Technology, 02-822 Warsaw, Poland

[2] Institute of High Pressure Physics PAS, 01-142 Warsaw, Poland

*Correspondence: maksimdub19f94@gmail.com

## Abstract

Terahertz transmittance spectra of plasmonic crystals based on two-dimensional electron gas in AlGaN/GaN heterostructures were studied in grating-gate and gateless plasmonic crystals as a function of lateral bias. The decrease of the plasmon resonance frequencies (redshift) with increase of the lateral current was observed for both types of structures. We show that the change of the electron concentration profile and Joule heating are the main phenomena responsible for the shift of plasma resonant frequency. These results are important for designing plasmonic resonances based filters, detectors, and emitters operating under voltage bias conditions.

Plasmons in two-dimensional (2D) systems are of great interest for over 50 years. The frequency of this kind of electrons' collective oscillations in AlGaN/GaN, AlGaAs/GaAs heterostructures, and Si inversion layers lies in the terahertz (THz) frequency band and can be tuned by geometry, electric field, temperature, etc (see ref.[[1]] and references therein). The analysis of the electron drift (lateral current) on the plasmon's properties started more than 40 years ago[2],[3],[4]. The effect of current on the plasmon properties is of special interest because it can potentially lead to amplification and generation at THz frequencies.

Already in the earliest work[3], it was shown that under conditions of the 2DEG drift, the amplification of the plasmons can be achieved. It was also shown that this effect can occur at a drift velocity significantly smaller than the plasmon velocity. Another mechanism of plasmon instability initiated by current in FETs was proposed in 1993[5]. Experimental observation of these effects and electrically driven THz emission from devices with plasma waves turned out to be a more challenging task. Indeed, there were a few publications on THz emission from plasma wave devices based on Si transistor and III-V-based high electron mobility devices [6,7,8,9,10]. However, in all cases, the emission signal was very weak and broadband. In some cases, reflectance of the surrounding background thermal radiation from the cold sample can be confused with the emission[11].

In all experiments attempting to observe the terahertz emission from the devices with plasma waves, DC or pulsed voltage was applied, and the emission signal was checked with some kind of the detector. However, this kind of the experiments does not allow to study the effect of the current (electron drift) on the properties of plasmons. Particularly, current should produce Doppler effect, which is manifested by the change of plasma resonant frequency. Current can increase temperature and affect electron concentration profile, which also may lead to the change of the plasmon frequency and plasmon resonance quality factor.

We are aware of only two publications where plasmon properties in AlGaAs/GaAs heterostructures were studied as a function of the drift current [12,13]. It was shown that with the current increase, the plasmon frequency decreases. This was explained by the Doppler effect but splitting of the plasmon frequency expected in the Doppler effect was not observed. The effects of temperature and lateral current on electron concentration profile were not considered in those publications.

Here, we studied experimentally the effect of current on plasmons in AlGaN/GaN heterostructures. We analyzed the transmittance spectra in large area (up to 6.25 mm$^2$)

AlGaN/GaN structures with a periodic charge density profile. This kind of structures, known as plasmonic crystals (PCs), were formed by periodic gates (Fig.1(a),(c)) and by periodic Ohmic contacts (Fig.1(b),(d)).

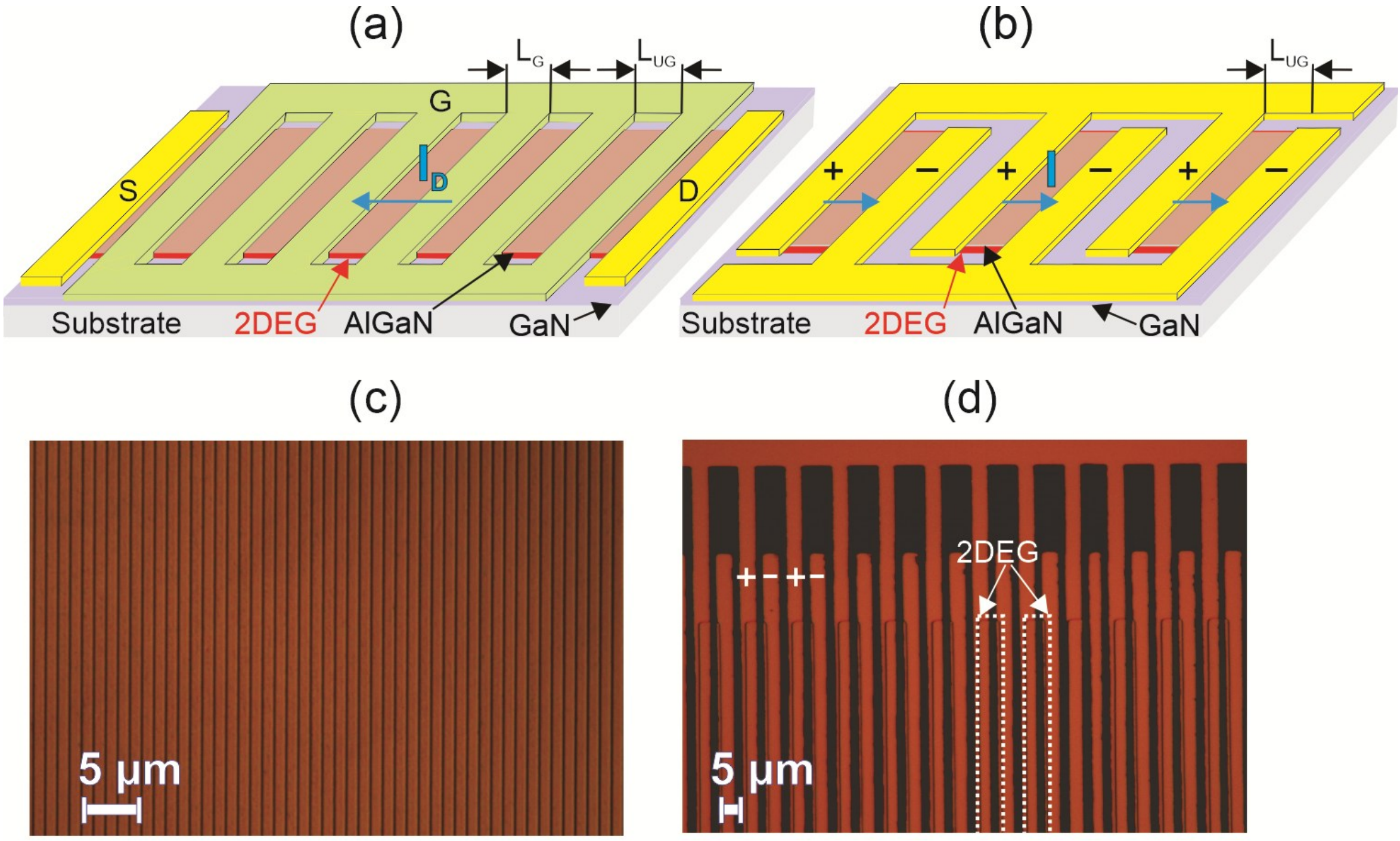

FIG. 1. (a)-(b) Schematics and (c)-(d) optical microscope images of the studied devices. (a),(c) Grating-gate FET. (b), (d) Gateless periodic structure.

Commercially available heterostructures from “SweGaN” (Linköping, Sweden[14]) were used to fabricate two types of large area PCs, schematically shown in Figs. 1(a)–1(b). Figures 1(c)-1(d) show the optical microscope images of corresponding PCs. The semiconductor stack consisted of a 2.4 nm GaN cap, a 20.5 nm Al0.25Ga0.75 N barrier, and a 255 nm GaN buffer grown directly on 62-nm-thick AlN nuclear layer on SiC substrate.

The PC in Fig. 1(a) represents the grating-gate structure fabricated in the configuration of the field effect transistor (FET) with source (S), drain (D), and gate (G) terminals. Figure 1(b) shows schematically the periodic gateless structure. The conducting 2D electron gas in these structures was removed in every second cavity by the reactive ion etching (RIE) (grey areas in Fig.

1(b)) in order to avoid the current flowing in the opposite directions in neighboring cavities when DC voltage is applied.

The laser writer system and Electron beam lithography (EBL) were used to define large and submicron elements, respectively. The Schottky barrier gates were fabricated using Ni/Au (150/350 Å) metal stack. For the ohmic contacts, the Ti/Au/Ni/Au (150/1000/400/500 Å) metal stacks were deposited and annealed at 800 °C in a nitrogen atmosphere for 1 minute (see ref.[1] for more details). The main parameters of the fabricated PCs are shown in Table I.

Table I. Main parameters of the studied PCs.

| sample ID | Cavity lengths, gated/ungated $L_G/L_{UG}$, μm | Number of cavities: gated/ungated | Active area, $mm^2$ | Total channel width, mm |
|---|---|---|---|---|
| S4 | 0/3 | 0/245 | 6.25 | 600 |
| S6 | 0/2 | 0/267 | 6.25 | 700 |
| S7 | 0.9/06 | 550/550 | 2.89 | 1.7 |

The transmission spectra of the plasmonic crystals were studied using a Fourier-transform infrared (FTIR) Vertex 80v, Bruker spectrometer (see ref. [15] for more details).

The structure shown in Fig. 1 (a) represents a FET with an exceptionally long total gate length of $L_{Gtotal}$=550×$L_G$≈0.5mm ($L_G$ is the individual gate finger length). When the gate voltage is zero and drain voltage is applied, the electric field in the channel increases until the drain voltage reaches the threshold voltage. In the typical AlGaN/GaN heterostructures, the electron concentration in the quantum well at the interface is of the order $10^{13}$ $cm^{-2}$. With the Schottky barrier gate and thickness of the AlGaN barrier layer of 25-30 nm, the threshold voltage ranges from $V_{th}$=-3V to $V_{th}$ =-4V. At drain voltages $V_D > |V_{th}|$, the transistor is in saturation, and the drain current is almost drain voltage independent. All extra voltage applied above the saturation voltage drops on the narrow part of the channel close to the drain. Indeed, this narrow region of the high electric field increases in length with the drain voltage increase. However, the estimate based on the approach developed in ref.[16] shows that it does not exceed a few microns in the length at all reasonable drain voltages, which is negligible in comparison with the total gate length of ~0.5 mm and source-to-drain distance of $L_{DS}$ ~1.7 mm. Since the current does not change above saturation, the electric field in the middle of the channel, $E_m$, does change as well and remains of the order $E \approx V_{th}/2L_{DS}$ [17]. For the typical threshold voltage $V_{th}$ = 4 V and $L_{DS}$ = 1.7 mm, the electric field is $E_m \approx 10$ V/cm. This electric field corresponds to the electron drift velocity, $v_D$, about three orders of magnitude smaller than the typical plasmon velocity of the order $10^8$ cm/s.

In general, the higher electron drift velocity can be achieved using the PC structure of another geometry, schematically shown in Fig. 1 (b). This is a gateless structure with interdigitated contacts. Although this kind of structure allows, in principle, to reach high drift velocity, unrealistically high total current is required. For example, to reach the velocity $v_D \sim 10^6$ cm/s in the S4 PC, the current $I \sim 100$ A is required. This is because of the huge total width of the structure of the order $W \approx 0.7$ m.

Nevertheless, we observed that the plasmon frequency can still be tuned by the drain voltage in the PCs shown in Fig. 1(a) and by the current in the PCs shown in Fig. 1(b). Figure 2(a) and 2(b) show the transmittance spectra for S7 PC (FET-like structure) and gateless S4 PC as a function of lateral bias at temperature $T$= 70 K. Minima in the transmission spectra correspond to the resonant frequency of the first and second harmonics of the plasmon resonances. The resonant frequency depends on the PC period, electron concentration, and other parameters like electron effective mass, dielectric permittivity, geometry[1,14], etc. In PCs such as S7 grating-gate FET, the interaction between the gated and ungated regions plays an important role in forming 2D delocalized plasmons[14]. The resonant frequency, $f_{DL}$, in this kind of PC depends on the electron concentration in both gated and ungated parts. This kind of plasmon oscillation is called the *delocalized mode*[14]. The gateless periodic structure S4 PC (Fig. 2(b)) does not contain gates and supports only one type of plasmonic cavity, defined between the ohmic contacts. Plasmon oscillations in this kind of PC belong to the *localized mode*[14]. The resonant frequency, $f_L$, in this kind of PC depends on the electron concentration between ohmic contacts and is bias voltage independent. Plasmon velocities, which define the plasmon frequency, are proportional to the square root of electron concentration for both gated and ungated cavities [1,14].

As seen in Fig.2 , in both PCs the plasmon frequency redshifts with the bias increase. When a drain voltage is applied to the transistor-like grating-gate structure (S7), the concentration in the channel under the gates becomes coordinate-dependent. At the drain voltage above saturation ($V_D > |V_{th}| \approx 3$V for S7 PC), the concentration at the drain side is virtually zero; the concentration at the source side is equal to the equilibrium channel concentration. Since the average concentration becomes smaller than that at zero bias, the delocalized plasmon frequency shifts to the lower frequencies. Increase of the drain voltage above the threshold up to ~ 6 V does not change the plasmon frequency. This is an expected result because extra drain voltage applied above the threshold (saturation) drops in the narrow region near the drain and therefore does not change the concentration profile.

As seen in Fig. 2(b), the plasmon frequency redshifts with the current increase in the gateless PC S4. Electron concentration between ohmic contacts in this gateless structure does not depend on current. The estimate for the frequency shift due to the Doppler effect at the highest current yields the value of the order $\Delta f = v_D/L_{UG} < 0.1$ GHz, which is negligible in comparison with the plasmon frequency.

The effect can be explained by Joule heating of the devices and the corresponding resonant frequency redshift [11,18].

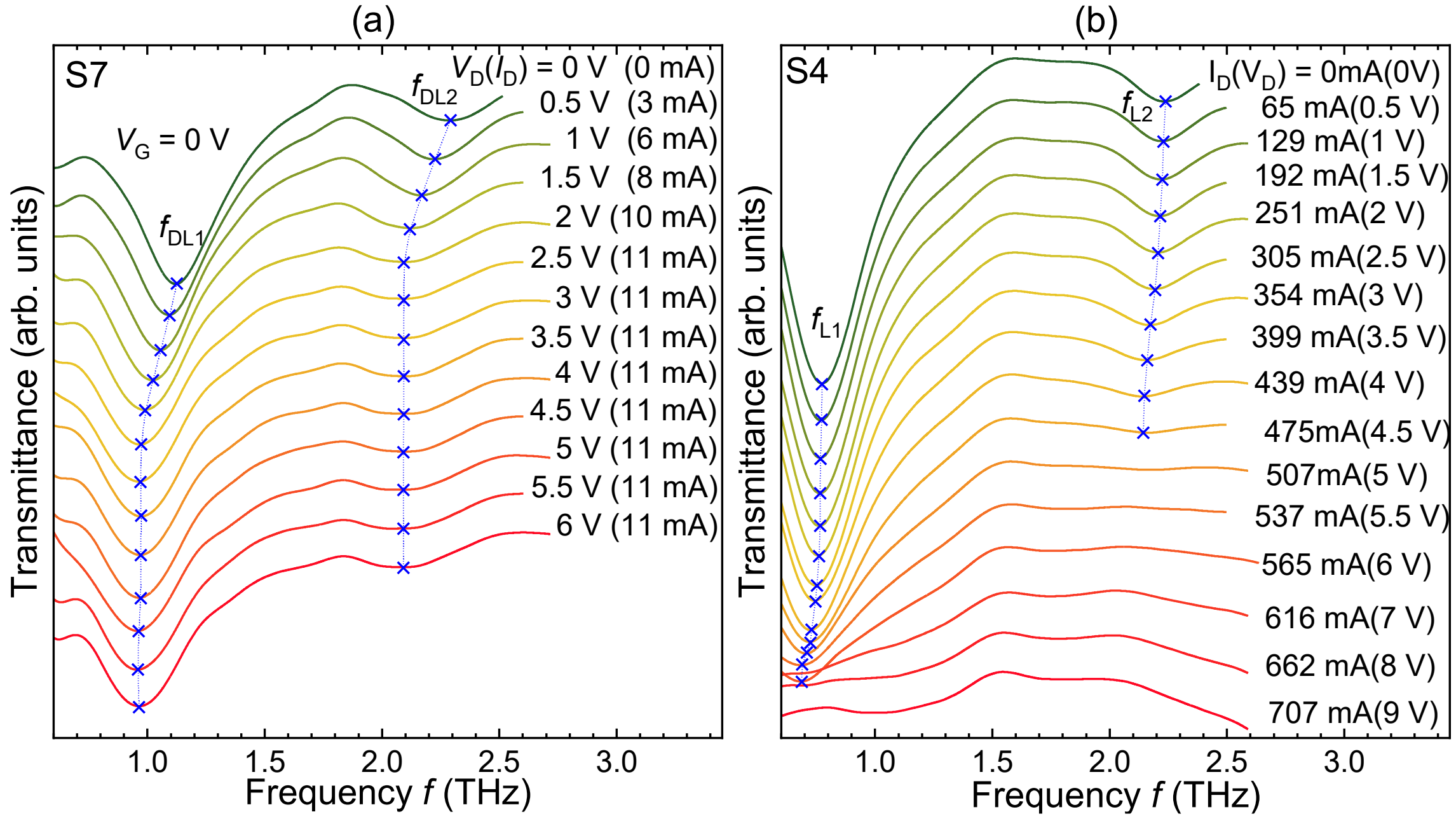

FIG. 2. (a) Transmittance spectra for the S7 PC at different drain voltages and (b) for the S4 PC at different currents at $T$= 70 K. Blue crosses show the position of the plasmon resonance frequencies. Dashed lines are guides for the eye.

To prove this we plotted in Fig. 3 the resonant frequency ($f_L$) as a function of the resistance measured at small bias at different temperatures (filled symbols) and the resonant frequency measured at different biases at 70 K (open symbols) for the S4 and S6 PCs (See Fig. S1 in the supplementary materials for S6 PCs transmittance spectra at different currents and temperatures). As seen, both dependencies coincide, indicating that the frequency change is due to the Joule heating. Therefore, we conclude that temperature increase due to the Joule heating plays an important role in the bias dependence of the plasmon frequency in these PCs. The temperature dependence of the plasmon properties in AlGaN/GaN PCs was studied in ref.[19,20,18]. It was shown that the temperature dependence of the electron effective mass and electron concentration are the main reasons for the plasmon resonant frequency temperature dependence [18].

Note that the Joule heating has little effect on delocalized plasmon properties of the grating-gate structure (PC S7, Fig.2(a)) because of a small dissipated power.

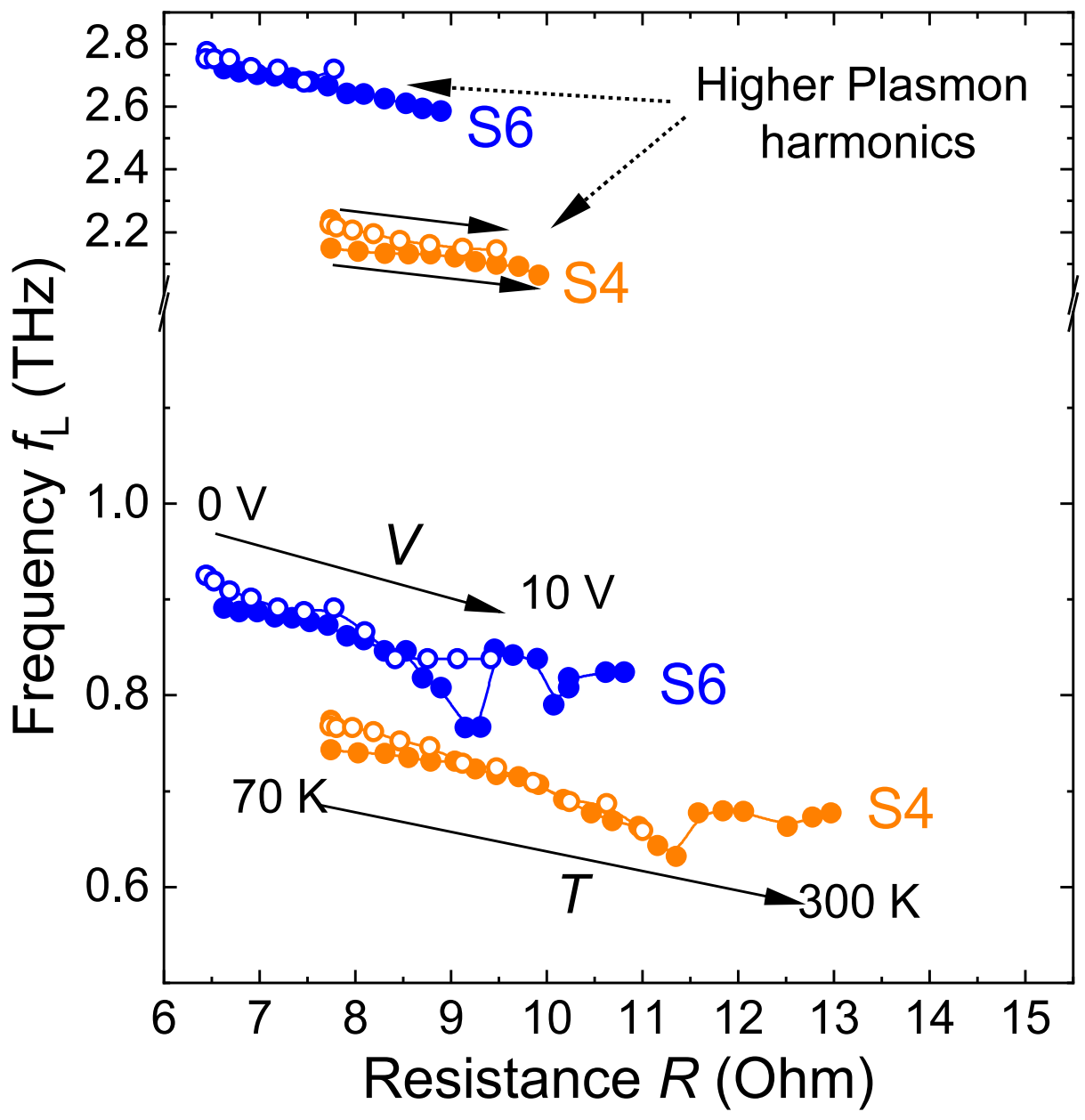

FIG. 3. Dependence of the plasmon resonance frequency on the PCs resistance changed by temperature (filled symbols) and by the bias (open symbols) for two PCs.

To conclude, the effect of the lateral bias on terahertz plasmons was studied in large-area AlGaN/GaN plasmonic crystals of two different types. Grating-gate PCs demonstrated the redshift of the resonant plasmon frequency with the drain voltage increase from zero to the threshold voltage. This effect is due to the change of the electron concentration profile between source and drain. The increase of the drain voltage above threshold does not alter the plasmon frequency because all extra voltage drops on the narrow region close the drain keeping the electron concentration profile for the majority of the structure unchanged.

In the gateless PCs, the plasmon frequency decreases with the current increase due to the Joule heating. The effect of the temperature on the plasmon properties was studied earlier, where it was shown that red shift of the plasmon frequency is caused by the temperature dependence of the electron effective mass and 2D electron concentration [18].

The effects of the Joule heating and dependence of the electron concentration profile on current are much stronger than expected Doppler effect, which probably cannot be observed in large area plasmonic crystals. Nevertheless, in small submicron devices, the Doppler effect should be essential. These results are important for designing and understanding the behavior of large-area PCs detectors, filters, and emitters that frequently operate under voltage bias conditions.

SUPPLEMENTARY MATERIAL

See the supplementary material for additional information and figures.

The work was supported by the European Union through the ERC-ADVANCED grant TERAPLASM (No. 101053716). Views and opinions expressed are, however, those of the author(s) only and do not necessarily reflect those of the European Union or the European Research Council Executive Agency. Neither the European Union nor the granting authority can be held responsible for them. We also acknowledge the support of "Center for Terahertz Research and Applications(CENTERA2)" project (FENG.02.01-IP.05-T004/23) carried out within the "International Research Agendas" program of the Foundation for Polish Science, co-financed by the European Union under European Funds for a Smart Economy Programme.

AUTHOR DECLARATIONS

Conflict of Interest

The authors declare no conflicts of interest.

Author Contributions

M. Dub: Conceptualization (lead); Data curation (lead); Investigation (lead); Methodology (equal); Project administration (lead); Writing– original draft (lead); Writing– review & editing (equal). P. Sai: Investigation (equal); Methodology (equal); Writing– review & editing (lead). Y. Ivonyak: Resources (equal); Investigation (equal). P. Prystawko: Resources (equal); Writing– review & editing (equal). W. Knap; Conceptualization (equal); Supervision (lead); Validation (lead); Funding acquisition (lead); Writing– review & editing (equal).S. Rumyantsev: Conceptualization (lead); Data curation (lead); Methodology (lead); Project administration (lead); Writing– original draft (lead); Writing– review & editing (equal); Supervision (equal); Validation (equal).

REFERENCES

1 P. Sai, V.V. Korotyeyev, M. Dub, M. Słowikowski, M. Filipiak, D.B. But, Y. Ivonyak, M. Sakowicz, Y. M. Lyaschuk, S.M. Kukhtaruk, G. Cywiński, and W. Knap, Phys. Rev. X **13** (4), 041003 (2023).

2 A.V. Chaplik, Surface Science Reports **5** (7), 289 (1985)..

3 A.V Chaplik, JETP Lett.(Engl. Transl.);(United States) **32** (8) (1980).

4 A.V. Chaplik, Solid state communications **65** (12), 1589 (1988).

5 M. Dyakonov and M. Shur, Physical Review Letters **71** (15), 2465 (1993).

6 W. Knap, J. Lusakowski, T. Parenty, S. Bollaert, A. Cappy, V.V. Popov, and M.S. Shur, Applied Physics Letters **84** (13), 2331 (2004).

7 W. Knap, F. TEPPE, N. DYAKONOVA, and A. El FATIMY, IEICE transactions on electronics **89** (7), 926 (2006).

8 Y.M. Meziani, H. Handa, W. Knap, T. Otsuji, E. Sano, V.V. Popov, G.M. Tsymbalov, D. Coquillat, and F. Teppe, Applied Physics Letters **92** (20) (2008).

9 T. Otsuji, Y.M. Meziani, T. Nishimura, T. Suemitsu, W. Knap, E. Sano, T. Asano, and V.V. Popov, Journal of Physics: Condensed Matter **20** (38), 384206 (2008).

10 V.A. Shalygin, M.D. Moldavskaya, M.Ya. Vinnichenko, K.V. Maremyanin, A.A. Artemyev, V. Yu. Panevin, L.E. Vorobjev, D.A. Firsov, V.V. Korotyeyev, and A.V. Sakharov, Journal of Applied Physics **126** (18), 183104 (2019).

11 M. Dub, D.B. But, P. Sai, Yu. Ivonyak, M. Słowikowski, M. Filipiak, G. Cywinski, W. Knap, and S. Rumyantsev, AIP Advances **13** (9) (2023).

[12] R.E. Tyson, R.J. Stuart, D.E. Bangert, R.J. Wilkinson, C.D. Ager, H.P. Hughes, C. Shearwood, D.G. Hasko, J. Frost, and D.A. Ritchie, Superlattices and microstructures **12** (3), 371 (1992).

[13] R.E. Tyson, R.J. Stuart, H.P. Hughes, J. Frost, D.A. Ritchie, G. Jones, and C. Shearwood, International journal of infrared and millimeter waves **14**, 1237 (1993).

[14] https://swegan.se.

[15] M. Dub, P. Sai, P. Prystawko, W. Knap, and S. Rumyantsev, Nanomaterials **14** (18), 1502 (2024).

[16] M. S. Shur, IEEE Transactions on Electron Devices **32** (1), 70 (1985).

[17] M. Shur, Inc., Englewood Cliffs, New Jersey, 680 (1990).

[18] M. Dub, P. Sai, D. Yavorskiy, Y. Ivonyak, P. Prystawko, R. Kucharski, G. Cywinski, W. Knap, and S. Rumyantsev, ArXiv preprint arXive (2025).

[19] D. Pashnev, V.V. Korotyeyev, J. Jorudas, T. Kaplas, V. Janonis, A. Urbanowicz, and I. Kašalynas, Applied Physics Letters **117** (16) (2020).

[20] V.V. Korotyeyev, V.A. Kochelap, V.V. Kaliuzhnyi, and A.E. Belyaev, Applied Physics Letters **120** (25) (2022).

# Supplementary materials

# Effect of lateral current on terahertz plasmons in AlGaN/GaN

M. Dub[1,2*], P. Sai[1,2], Y. Ivonyak[1,2], P. Prystawko[2], W. Knap[1,2], S. Rumyantsev[2]

[1]CENTERA, CEZAMAT, Warsaw University of Technology, 02-822 Warsaw, Poland

[2] Institute of High Pressure Physics PAS, 01-142 Warsaw, Poland

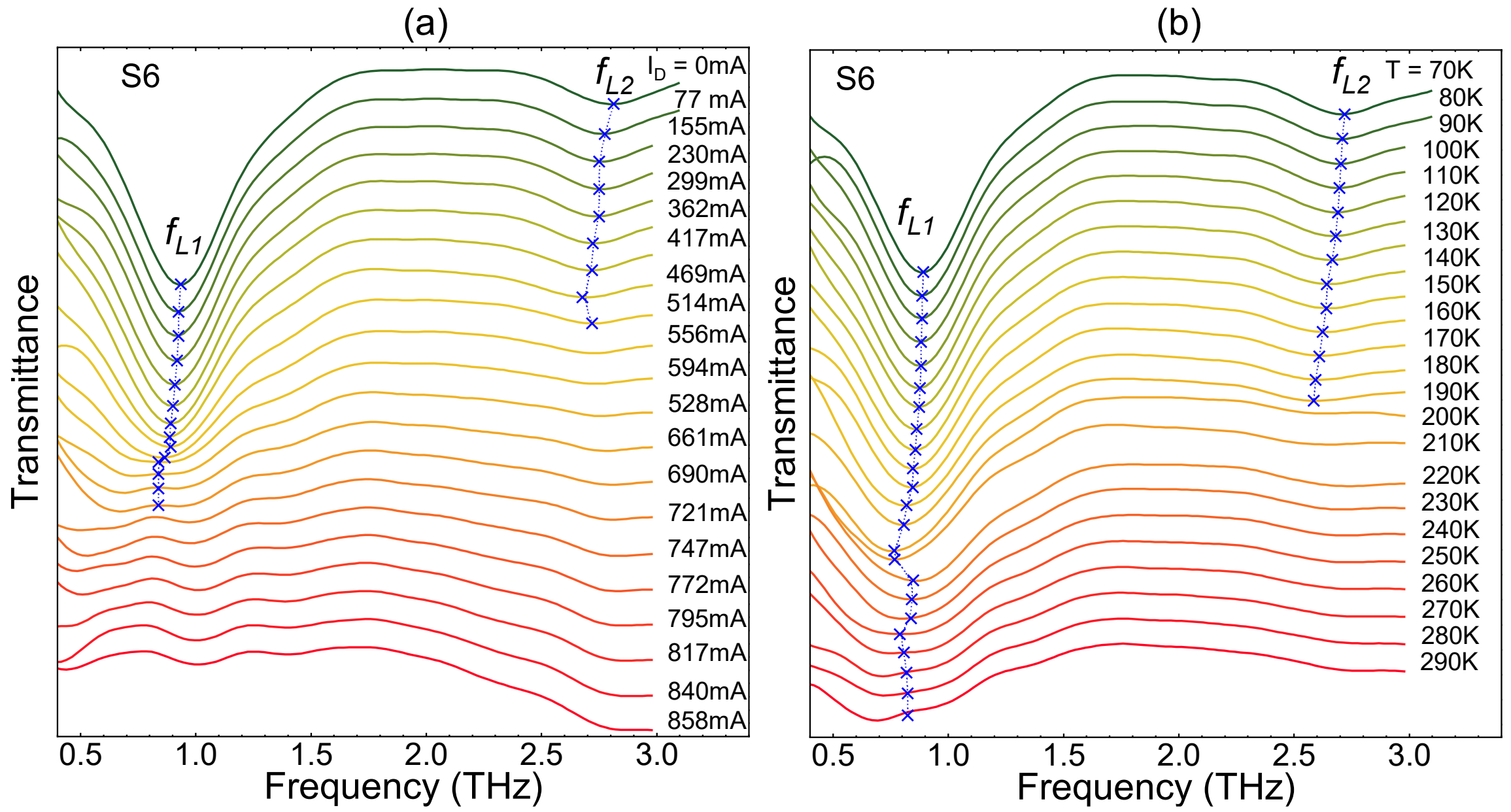

FIG. S1. Transmittance spectra for the S6 PC at different currents (a) and temperatures (b). Crosses show the position of the plasmon resonance frequencies. Dashed lines are guides for eyes.